# VARIABLE POLARIZATION CONTROL IN FREE-ELECTRON LASERS

H.P. Freund[1,2] and P.J.M. van der Slot[3]

[1]Department of Electrical and Computer Engineering, University of New Mexico, Albuquerque, New Mexico, USA
[2]NOVA Physical Science and Simulations, Vienna, Virginia, USA
[3]Laser Physics and Nonlinear Optics, Mesa[+] Institute for Nanotechnology, University of Twente, Enschede, the Netherlands

Free-electron lasers (FELs) over virtually the entire electromagnetic spectrum from microwaves through ultraviolet through hard x-rays that are either seeded or start from noise (SASE). FELs can produce a variety of different optical polarizations of the output radiation ranging from linear through elliptic to circular polarization depending upon the characteristics of the undulators used. For example, x-ray FELs employ long undulator lines composed of a series of relatively short undulators. Most such FELs use linearly polarized undulators due to the ease of manufacture and tuning; hence the optical output is linearly polarized. However, elliptic or circular polarizations may be achieved by varying the orientation of the undulators along the line. Alternately, APPLE-II or Delta undulator designs allow for producing undulating magnetic fields with arbitrary polarizations. In this paper, we present a three-dimensional, time-dependent formulation that self-consistently models two independent optical polarizations and, therefore, completely treats any given sequence or combination of undulator, undulator imperfections, and undulator degradation There are two principal characteristics of the formulation that underpin this capability. In the first place, the particle trajectories are integrated by means of the full Lorentz force equations using three-dimensional analytic models of the undulators fields. This permits an accurate model of the interaction of the electrons with a large variety of undulator fields and orientations. In the second place, the formulation allows the electrons to couple simultaneously to two independent polarizations of the electromagnetic wave and, therefore, allows the optical polarization to evolve self-consistently along the undulator line. After first describing the numerical model, we proceed to give some examples using different undulator configurations that are of interest.

## I. INTRODUCTION

Free-electron lasers (FELs) have been an intense area of research since the 1970s, and the development of fourth generation x-ray light sources [1-5] keep the field fresh. Intensive work is ongoing into new FEL-based light sources that probe ever shorter wavelengths with a variety of configurations. In particular, there is interest in the generation of a variety of different optical polarizations. These x-ray FELs employ long undulator lines composed of a sequence of relatively short undulators separated by gaps that may contain focusing magnets (typically quadrupoles), phase shifters (dipoles) and various diagnostics. The most common undulator type in use is linearly polarized and this means that the optical output from this configuration is also linearly polarized. However, undulator imperfections and possible undulator degradation may result in a polarization state different from the purely linear. Furthermore, there is interest in generating a variety of different polarizations. The most straightforward way to accomplish this is to use a line of linearly polarized undulators where the orientations of the undulators vary in some fashion along the undulator line. For example, the undulator line might be configured consisting of a first stage with a given undulator polarization followed by a second stage with a different undulator polarization that functions as an afterburner [6-9]. Since the interaction is close to saturation in the first stage, the electron beam is pre-conditioned for strong excitation in the afterburner and this can generate elliptically polarized output. A configuration like this has been implemented on the Linac Coherent Light Source (LCLS) at the Stanford Linear Accelerator Center (SLAC) using the so-called Delta undulator [10] as an afterburner [11]. Alternately, the APPLE-II [12] undulator design is configurable to produce any type of undulator field from linearly polarized through elliptic polarization to a helically polarized field and this can be used for the complete undulator line [13] or as an afterburner. In view of the growing interest in novel undulator configurations, it is important to develop a numerical formulation that can self-consistently model the evolution of the polarization of the light through an arbitrary sequence of undulators.

The bulk of existent FEL codes employ the slowly-varying envelope approximation (SVEA) [14-20] and use field models with fixed polarizations to describe the interaction with undulators with linear, helical or elliptical polarizations. An exception to this is the particle-in cell code PUFFIN which can treat arbitrary polarizations [21].

In this paper, we describe a three-dimensional, time-dependent nonlinear SVEA formulation of the interaction in FELs that self-consistently models arbitrary optical polarizations due to varying undulator configurations, imperfections in the magnetic field, possible undulator degradation, or any combination of these effects. We apply this formulation to the study of variable polarizations produced by an APPLE-II undulator line. Particle dynamics are treated using the full Newton-Lorentz force equations to track the particles through the optical and magnetostatic fields (including undulators, quadrupoles and dipole phaser shifters). This permits the treatment of particle dynamics in the native fields of any arbitrary configuration. The optical field is described by a superposition of Gaussian modes, and the formulation tracks the particles and fields as they propagate along the undulator line from the start-up through the (linear) exponential growth regime and into the nonlinear post-saturation state. The formulation includes three-dimensional descriptions of linearly polarized, helically polarized, and elliptically polarized undulators including the fringing fields associated with the entry/exit transition regions. Analytic magnetostatic field models for quadrupoles and dipoles are also included. The optical field is described by a superposition of Gaussian modes, and the formulation self-consistently tracks the particles and optical field through the undulator line, which may include gaps, quadrupoles and dipoles from the start-up through the (linear) exponential growth regime and into the post-saturation state.

As such, the present formulation follows that described previously for the magnetostatic fields and particle tracking [17,18]. The difference comes in with the treatment of the electromagnetic field. Instead of prescribing a fixed phasor polarization for the optical field, the new formulation introduces two perpendicular field components in the $x$- and $y$-directions with the $z$-direction along the axis of symmetry of the undulator line, that simultaneously interacts with the electrons. The formulation allows the polarization of the light to dynamically evolve upon propagation along the undulator line. Both formulations use the Slowly-Varying Envelope Analyses (SVEA) where the optical field is represented by a slowly-varying amplitude and phase in addition to a rapid sinusoidal oscillation. The field equations are then averaged over the rapid sinusoidal time scale and, thereby, reduced to equations describing the evolution of the slowly-varying amplitude and phase. In the new formulation there are now twice as many field equations to be integrated as previously. Fortunately, this presents a relatively small additional computation load on modern computers where vectorization and multiple processors can be employed.

The organization of the paper is as follows. The formulation is described in Section II. Examples are given in Sec. III. A summary and discussion is given in Section IV.

**II. THE NUMERICAL FORMULATION**

To summarize, the numerical formulation describes the particles and fields in three spatial dimensions and includes time dependence as well. Electron trajectories are integrated using the complete Newton-Lorentz force equations. No wiggler-averaged-orbit approximation is made. The magnetostatic fields are specified by analytical functions for a variety of undulator models (such as planar, elliptical, or helical representations), quadrupoles, and dipoles. These magnetic field elements can be placed in arbitrary sequences to specify a variety of different transport lines. As such, field configurations can be set up for single or multiple undulator segments with quadrupoles either placed between the undulators or superimposed upon the undulators to create a FODO lattice. Dipole chicanes can also be placed between the undulators to model various optical klystron and/or high-gain harmonic generation (HGHG) configurations or phase shifters. The field models and a statement of the dynamical orbit equations has been described previously [18,19]. As before, we still refer to the new simulation code as MINERVA.

The electromagnetic field is described by a modal expansion using Gaussian optical modes. In the earlier formulation [18,19], we used the Gauss-Hermite modes for the simulation of planar undulators, while Gauss-Laguerre modes are used for elliptical or helical undulators. The present model uses only the Gauss-Hermite modes; however, comparison of helical wiggler simulations with both formulations shows near-identical results using the two Gaussian mode representations; hence, there is no loss of generality using only the Gauss-Hermite modes.

Since, the electromagnetic field representations are also used in integrating the electron trajectories, harmonic motions and interactions are included self-consistently. The same integration engine is used within the undulator(s) as in the gaps, quadrupoles, and dipoles, so that the phase of the optical field relative to the electrons is determined self-consistently when propagating the particles and fields in the gaps between the undulators. The slippage of the electromagnetic field relative to the electrons is also included self-consistently in the gaps.

The representation of the Gauss-Hermite modes in present use are given by

$$\delta \mathbf{A}(\mathbf{x},t) = \hat{e}_x \sum_{\substack{l,n=0 \\ h=1}}^{\infty} e_{l,n,h}^{(x)} \left( \delta A_{l,n,h}^{(1,x)} \sin \varphi_h^{(x)} + \delta A_{l,n,h}^{(2,x)} \cos \varphi_h^{(x)} \right)$$

$$+ \hat{e}_y \sum_{\substack{l,n=0 \\ h=1}}^{\infty} e_{l,n,h}^{(y)} \left( \delta A_{l,n,h}^{(1,y)} \sin \varphi_h^{(y)} + \delta A_{l,n,h}^{(2,y)} \cos \varphi_h^{(y)} \right), \quad (1)$$

where the indices ($l,n$) describe the transverse mode structure, the index $h$ is the harmonic number, $\delta A_{l,n,h}$, are the mode amplitudes.

$$e_{l,n,h}^{(j)} = \frac{w_{0,h}^{(j)}}{w_h^{(j)}} \exp\left(-r^2 / w_h^{(j)2}\right) H_l\left(\zeta_x^{(j)}\right) H_n\left(\zeta_y^{(j)}\right), \quad (2)$$

describes the transverse mode structure ($j = x,y$) where $H_l$ and $H_n$ are the Hermite polynomials, $\zeta_x^{(j)} = \sqrt{2}x / w_h^{(j)}$, $\zeta_y^{(j)} = \sqrt{2}y / w_h^{(j)}$, $w_{0,h}^{(j)}$ and $w_h^{(j)}$ denote the waist size and spot size of the $h^{\text{th}}$ harmonic in the $x$- and $y$-directions respectively. The phase is

$$\varphi_h^{(j)} = h\left(k_0 z - \omega t\right) + \alpha_h^{(j)} \left(\frac{r}{w_h^{(j)}}\right)^2, \quad (3)$$

where $k_0 = \omega/c$, $\alpha_h^{(j)}$ denotes the curvature of the phase front of the $h^{\text{th}}$ harmonic. The mode amplitudes, the spot sizes and the curvature of the phase front are assumed to be a slowly-varying function of ($z,t$).

The total power carried in each mode, $P_{l,n,h}$, is given by integration of the Poynting vectors over the cross section. This is given by

$$P_{l,n,h} = \frac{m_e^2 c^5}{8e^2} 2^{l+n-1} l!n!$$
$$\times \sum_{j=x,y} \left(k_0 w_{0,h}^{(j)}\right)^2 \left(\delta a_{l,n,h}^{(1,j)2} + \delta a_{l,n,h}^{(2,j)2}\right), \quad (4)$$

for the Gauss-Hermite modes in both the $x$- and $y$-directions, where $\delta a_{l,n,h}^{(1,2)} (= e\delta A_{l,n,h}^{(1,2)}/m_e c^2)$ is the normalized field amplitude, and $m_e^2 c^5/8e^2 \cong 1.089$ GW.

The dynamical equations for the fields employ the source-dependent expansion (SDE) [22] which is an adaptive eigenmode algorithm in which the evolution of the spot size and curvature are determined self-consistently in terms of the interaction with the electron beam. As such, the dynamical equations for the fields are of the form

$$\frac{d}{dz}\begin{pmatrix}\delta a_{l,n,h}^{(1,j)} \\ \delta a_{l,n,h}^{(2,j)}\end{pmatrix} + K_{l,n,h}^{(j)}\begin{pmatrix}\delta a_{l,n,h}^{(2,j)} \\ -\delta a_{l,n,h}^{(1,j)}\end{pmatrix} = \begin{pmatrix}S_{l,n,h}^{(1,j)} \\ S_{l,n,h}^{(2,j)}\end{pmatrix}, \quad (5)$$

where $S_{l,n,h}^{(i,j)}$ are the source terms and $i = (1,2)$ and $j = (x,y)$ denoting the $x$- and $y$-directions,

$$\frac{d}{dz} = \frac{\partial}{\partial z} + \frac{1}{c}\frac{\partial}{\partial t} \quad (6)$$

is the convective derivative, and

$$K_{l,n,h}^{(j)} = F_{l,n}\left(\frac{\alpha_h^{(j)}}{w_h^{(j)}}\frac{dw_h^{(j)}}{dz} - \frac{1}{2}\frac{d\alpha_h^{(j)}}{dz} - \frac{1+\alpha_h^{(j)2}}{k_0 w_h^{(j)2}}\right), \quad (7)$$

for $F_{l,n} = 1 + l + n$. The source terms are given by

$$\begin{pmatrix}S_{l,n,h}^{(1,j)} \\ S_{l,n,h}^{(2,j)}\end{pmatrix} = \frac{\omega_b^2}{k_0 c^2}\frac{1}{2^{l+n-1}\pi}\frac{1}{l!n!}\frac{1}{w_{0,h}^{(j)2}}$$
$$\times \left\langle e_{l,n,h}^{(j)}\frac{\upsilon_j}{|\upsilon_z|}\begin{pmatrix}-\cos\varphi_h^{(j)} \\ \sin\varphi_h^{(j)}\end{pmatrix}\right\rangle, \quad (8)$$

where $\omega_b$ is the beam plasma frequency, and $\langle(...)\rangle$ denotes an average over the initial particle distribution and $\upsilon_j$ ($j = x, y$) are the Cartesian components of the velocity of an electron with $z$ taken along the axis of the undulator. A uniform distribution in initial phase and a Gaussian distribution in coordinate and momentum space is assumed and the particle average follows that used in the previous formulation [17,18]

$$\langle(...)\rangle = \int_0^{2\pi}\frac{d\psi_0}{2\pi}\int_1^{\infty}\frac{d\gamma_0}{\sqrt{\pi/2}\Delta\gamma}\frac{\exp\left[-(\gamma_0-\gamma_{avg})^2/2\Delta\gamma^2\right]}{1+\mathrm{erf}\left(\gamma_{avg}/\sqrt{2}\Delta\gamma\right)}$$
$$\times \iint\frac{dx_0 dy_0}{2\pi\sigma_r^2}\iint\frac{dp_{x0}dp_{y0}}{2\pi\sigma_p^2}\exp\left(-r^2/2\sigma_r^2 - p_{\perp 0}^2/2\sigma_p^2\right)$$
$$\times (...), \quad (9)$$

where $\gamma_{avg}$ and $\Delta\gamma$ denote the average energy and energy spread, and $\sigma_r$ and $\sigma_p$ describe the initial transverse phase space. The macro-electrons in the simulation are set by Gaussian quadrature. Once the initial phase space is generated, the coordinates and momenta for each macro-electron is tracked by the Newton-Lorentz equations [18,19] along with the optical field equations. The evolution of the spot size and curvature are governed by SDE equations that reflect the strength of the interaction. Using the SDE [22], the dynamical equations for the spot size and curvature are written in the form

$$\frac{dw_h^{(j)}}{dz} = \frac{2\alpha_h^{(j)}}{hk_0 w_h^{(j)}} - w_h^{(j)}Y_h^{(j)}, \quad (10)$$

$$\frac{1}{2}\frac{d\alpha_h^{(j)}}{dz} = \frac{1+\alpha_h^{(j)2}}{hk_0 w_h^{(x,y)2}} - X_h^{(j)} - \alpha_h^{(j)}Y_h^{(j)}, \quad (11)$$

where the source terms are defined as

$$X_h^{(j)} = \frac{2}{\delta a_{0,0,h}^{(j)\,2}}\left[\left(S_{2,0,h}^{(1,j)} + S_{0,2,h}^{(1,j)}\right)\delta a_{0,0,h}^{(1,j)}\right.$$
$$\left. - \left(S_{2,0,h}^{(2,j)} + S_{0,2,h}^{(2,j)}\right)\delta a_{0,0,h}^{(2,j)}\right], \quad (12)$$

$$Y_h^{(j)} = -\frac{2}{\delta a_{0,0,h}^{(j)\,2}}\left[\left(S_{2,0,h}^{(1,j)} + S_{0,2,h}^{(1,j)}\right)\delta a_{0,0,h}^{(1,j)}\right.$$
$$\left. + \left(S_{2,0,h}^{(2,j)} + S_{0,2,h}^{(2,j)}\right)\delta a_{0,0,h}^{(2,j)}\right], \quad (13)$$

where $\delta a_{0,0,h}^{(j)\,2} = \delta a_{0,0,h}^{(1,j)2} + \delta a_{0,0,h}^{(2,j)2}$. We remark that in the absence of an electron beam $X_h^{(x,y)} = Y_h^{(x,y)} = 0$ and vacuum diffraction is recovered.

The total number of equations in each simulation is

$$N_{equations} = N_{slices}\left[6N_{particles} + 4(N_{modes} + N_{harmonics})\right], \quad (14)$$

where $N_{slices}$ is the number of slices in the simulation, and for each slice, $N_{particles}$ is the number of particles, $N_{modes}$ is the number of modes in all the harmonics, and $N_{harmonics}$ is the number of harmonics. The coupled electron and optical field equations are integrated using a 4th order Runge-Kutta algorithm where small steps are used to resolve the wiggler motion (typically 20 – 30 steps per undulator period). The Runge-Kutta algorithm allows the step size to change "on the fly" so different steps can be used in quadrupoles, dipoles, or drift spaces between magnetic elements so this imposes no limitation on the placement of components along the electron beam path.

We rely on the Stokes parameters ($s_0$, $s_1$, $s_2$, $s_3$) to characterize the polarization of the optical field [23]. Further, we determine the eccentricity of the polarization ellipse as well as the orientation of the semi-major axis of the ellipse with respect to the $x$-axis [23] in case of elliptically polarized light, as illustrated in Fig. 1. If we denote the angle between the semi-major axis of the ellipse and the $x$-axis as $\theta$, then this angle is determined by

$$\theta = \frac{1}{2}\tan^{-1}\left(\frac{s_2}{s_1}\right). \quad (15)$$

The eccentricity, $\varepsilon$, of the ellipse is given by

$$\varepsilon = \sqrt{1 - \tan^2\chi}, \quad (16)$$

where

$$\chi = \frac{1}{2}\sin^{-1}\left(\frac{s_3}{s_0}\right). \quad (17)$$

The eccentricity of a purely circularly polarized optical field is zero while that for a purely linearly polarized optical field is unity. Elliptical polarizations are characterized by eccentricities between these extrema. Measurements of the polarization in FELs have been characterized by the Stokes parameters [24].

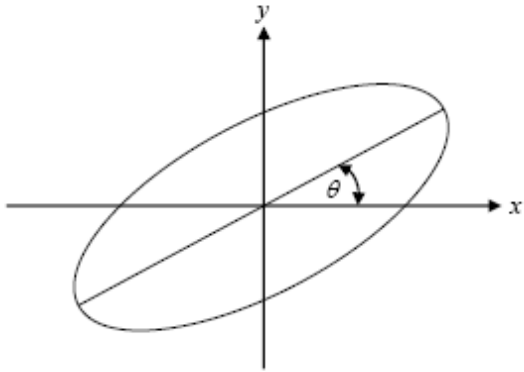

Fig. 1: Illustration of the ellipse formed by the polarization of the optical field.

At the present time, we have implemented a preliminary diagnostic for the Stokes parameters and the eccentricity based upon the on-axis field of the fundamental TEM$_{00}$ mode. A more complete diagnostic is under development.

## III. NUMERICAL SIMULATIONS

In this section we consider three different configurations. The first is a comparison with experimental observation from the "Sorgente Pulsata ed Amplificata di Radiazione Coerente" (SPARC) experiment which is a planar undulator-based SASE FEL located at ENEA Frascati [25] and which serves as validation of the numerical formulation. This experiment has been amply studied in simulation [18,19,25,26] with a variety of simulation codes. The fundamental parameters of the SPARC experiment are used for all the subsequent configurations, but with variations. The planar undulators and the quadrupole focusing lattice used in the experiment are replaced by a helical undulator in the second configuration. In this case, the electrons are matched into the weak focusing helical undulator. Here we compare the new formulation with Guass-Hermite modes and the dynamic evolution of the polarization with the old formulation that uses a prescribed circular polatization with Gauss-Laguerre modes. In the third configuration, we replace the planar undulators used in the experiment with Apple-II undulators which allows us to vary the polarization of the undulator line. Again, we compare the new formulation with the old formulation which uses a fixed elliptical polarization with Gauss-Laguerre modes.

| Electron Beam | |
|---|---|
| Energy | 151.9 MeV |
| Bunch Charge | 450 pC |
| Bunch Duration (rms) | 2.83 ps |
| $x$-Emittance | 2.5 mm-mrad |
| $y$-Emittance | 2.9 mm-mrad |
| rms Energy Spread | 0.02% |
| rms Size ($x$) | 132 microns |
| $\alpha_x$ | 0.938 |
| rms Size ($y$) | 75 microns |
| $\alpha_y$ | -0.705 |
| **Undulators** | 6 segments |
| Period | 2.8 cm |
| Length | 77 Periods |
| Amplitude | 7.8796 kG |
| $K_{rms}$ | 1.457 |
| Gap Length | 0.40 m |
| **Quadrupoles** | Centered in Gaps |
| Length | 5.3 cm |
| Field Gradient | 0.9 kG/cm |

Table 1: Parameters of the SPARC FEL experiment.

### A. The SPARC Experiment

In this section, simulations using the old and new formulations are compared corresponding to the SPARC experiment for which there are experimental measurements to anchor the two formulations. The experimental parameters of SPARC are summarized in Table 1. Pulse energies were measured in the gaps between the undulator segments. It was observed that the bunch charge was insufficient to reach saturation. In the simulation, the electron bunch was modeled using a parabolic profile with a peak current of 53 A and a width of 12.67 ps at its base. The simulation also assumed that each undulator segment had one period entrance and exit taper and that the electron beam was matched to the undulator/focusing lattice. For the parameters of table 1, the resonant wavelength is at 491.5 nm.

A comparison of the evolution of the pulse energy as found in simulation with the old (shown by black exes) formulation, the new formulation (blue) and as measured in the experiment (red markers) is shown in Fig. 2, where the simulations include an average taken over 20 simulation runs with different noise seeds which yields convergence to better than 5%. The pulse energy was measured in the gaps between the undulators, and the results are for a sequence of shots (data courtesy of L. Giannessi). Agreement between the two formulations is excellent and between the simulations and the measured performance is also quite good. Energy conservation in both the old and new formulations is maintained to within better than one part in $10^3$.

A comparison between the evolution of the relative linewidth as determined from simulations with the old formulation (black exes), the new formulation (blue) and by measurement (data courtesy of L. Giannessi) is shown in Fig. 3, where the two results of the two formulations are very close. Agreement between the simulations and the measured linewidth is within about 35% after 15 m. As shown, the predicted linewidths are in substantial agreement with the measurements.

The planar undulators used in the simulation were oriented in such a way that the wiggle-motion was aligned with the *x*-axis. As such, we expect that the optical field generated in simulation will be linearly polarized along the *x*-axis as well. For this case, the old formulation assumes the light to be polarized in the *x*-direction. In contrast, the polarization of the optical field will evolve self-consistently in the new formulation and and the on-axis Stokes parameters at the end of the undulator line yield $\theta = 0$ with an eccentricity $\varepsilon = 1$ as expected.

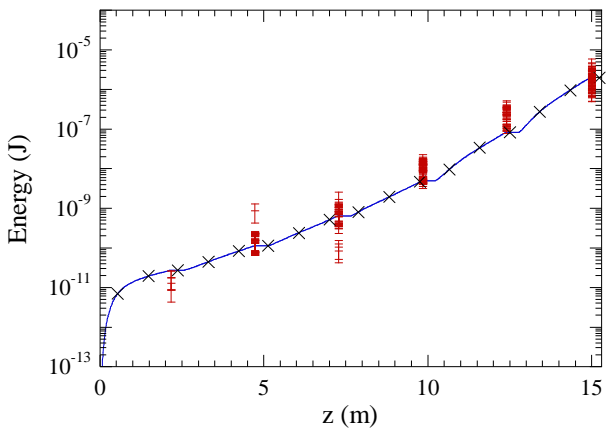

Fig. 2: Comparison of simulated and measured pulse energies versus distance (data courtesy of L. Giannessi). The new formulation is shown in blue, the old [17,18] is shown with black exes, and the measurements are denoted by red markers.

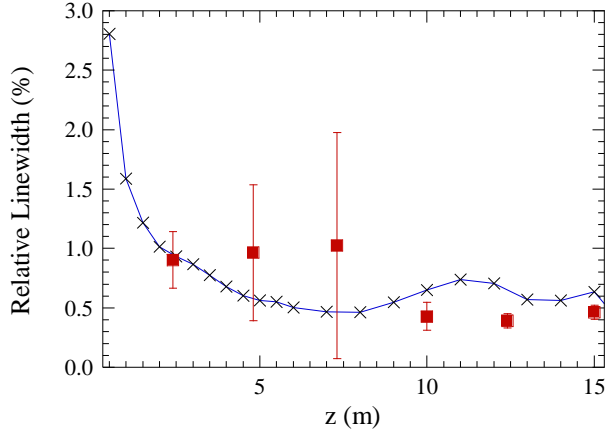

Fig. 3: Comparison of the measured relative linewidth in red (data courtesy of L. Giannessi) with that found in simulation. The new formulation is shown in blue, the old [17,18] is shown with black exes, and the measurements are denoted by the red markers.

**B. Helical Undulator Simulation**

For the helical undulator case, we employ the same electron beam parameters used for the simulation of the SPARC experiment but replace the undulator/quadrupole line with a long helical undulator. In order to preserve the resonance at 491.5 nm, the peak on-axis field of the helical undulator is reduced to 5.717 kG. In addition, the Twiss parameters are chosen to match the beam into the weak-focusing helical undulator [18,19]. As in the case of the SPARC experiment, we simulate SASE for this configuration but only use a single (and identical) noise seed for both formulations which means that the initial phase spaces in each simulation will be identical.

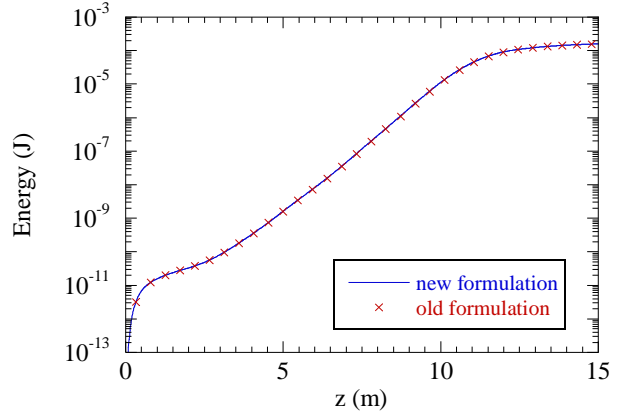

Fig. 4: Comparison of the evolution of the pulse energy using the present (new) and old [17,18] formulations.

Since there is no experiment which we can use to validate the formulation, we compare the present formulation with the results of simulation with the preceding model used in MINERVA [18,19], and we remark that the earlier formulation used a circularly polarized optical field representation based upon a superposition of Gauss-Laguerre modes in conjunction with a helical undulator. In contrast, the present formulation allows for the polarization of the optical field to vary based upon the electron motion induced by the undulator and is based upon a superposition of Gauss-Hermite modes. Since both the Gauss-Laguerre and Gauss-Hermite modes constitute complete sets, this difference should not be important as long as a sufficient number of modes is included.

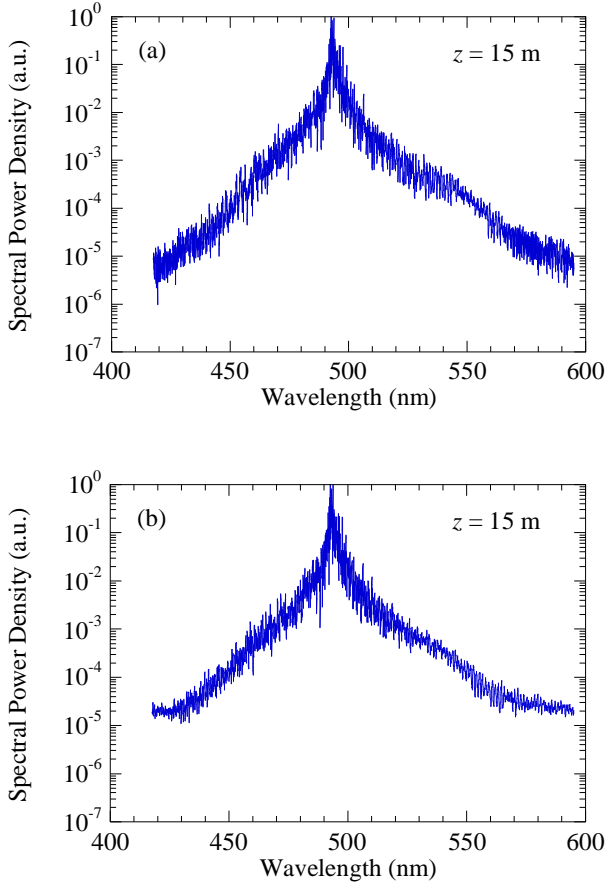

Fig. 5: Spectra found at the undulator exit using the old (a) and new (b) formulations.

Simulations of this configuration have been conducted with the earlier (old) formulation using 15 Gauss-Laguerre modes and with the present formulation using 12 Gauss-Hermite modes with 1600 temporal slices in each case. The evolution of the pulse energy with distance along the undulator is shown in Fig. 4, and we observe that excellent agreement is found between the two simulations. As in the simulations for the SPARC experiment, energy conservation is maintained to within better than one part in $10^3$.

The spectra produced by the two formulations is also in good agreement. This is evident in Fig. 5 where we plot the spectra at the undulator exit as found by the old (a) and new (b) formulations.

## C. The Apple-II Undulator

We use an analytic representation for the Apple-II undulator model [9,18,19] that is formed by the superposition of two flat-pole-faced planar undulators that are oriented perpendicularly to each other and shifted in phase ($\phi$). This takes the form

$$\mathbf{B}_w(\mathbf{x}) = B_w(z)\hat{e}_x \left( \sin(k_w z + \phi) - \frac{\cos(k_w z + \phi)}{k_w B_w} \frac{dB_w}{dz} \right) \times \cosh k_w x$$

$$+ B_w(z)\hat{e}_y \left( \sin k_w z - \frac{\cos k_w z}{k_w B_w} \frac{dB_w}{dz} \right) \cosh k_w y$$

$$+ B_w(z)\hat{e}_z \left[ \sinh k_w y \cos(k_w z + \phi) + \sinh k_w x \cos k_w z \right], \quad (18)$$

We define an ellipticity, $u_e$, related to the phase shift

$$u_e = \frac{1 - |\cos \phi|}{1 + |\cos \phi|}, \quad (19)$$

when $0 \le \phi \le \pi$. Note that with this definition, the ellipticity of the undulator is not the same as the eccentricity of the optical field since a linearly (helically) polarized undulator field is characterized by $u_e = 0$ (1).

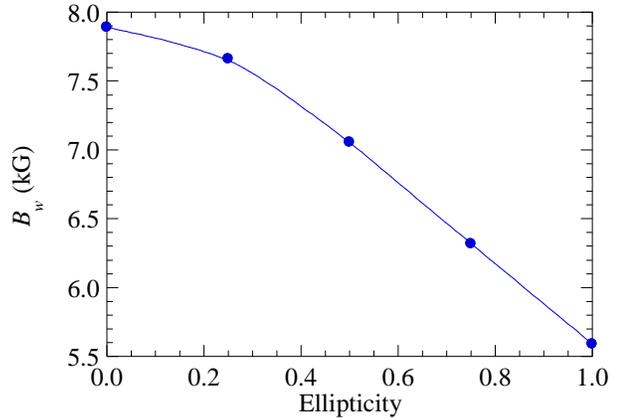

Fig. 6: Variation in the on-axis undulator field for various degrees of ellipticity, $u_e$, required for resonance at 491.5 nm.

We now consider simulations where we replace the linearly polarized undulators in the SPARC experiment with Apple-II undulators of the same lengths, and consider five different choices for the ellipticity: $u_e = 0.00, 0.25, 0.50, 0.75,$ and $1.00$ corresponding to a linearly polarized undulator field ($u_e = 0.00$), elliptically polarized undulator fields, and a helically polarized undulator field ($u_e = 1.00$) respectively. In order to ensure that the resonance remains at 491.5 nm, the on-axis field amplitudes for these cases were found to be 7.8796 kG, 7.6600 kG, 7.0577 kG, 6.3150 kG, and 5.5817 kG respectively as shown in Fig. 6 and as determined by [27]

$$\lambda_{res} = \frac{\lambda_w}{2\gamma^2}\left[1+\left(1+u_e^2\right)\frac{K^2}{2}\right], \quad (20)$$

where $\lambda_{res}$ is the resonant wavelength, and $K$ ($= eB_w/m_e c^2 k_w$) is the usual undulator strength parameter.

In order to study the interactions for these five cases out to saturation, and since the bunch charge was insufficient to reach saturation after 15 meters of undulators, we extend the undulator/quadrupole line to reach 28 meters in total length. We also note that the quadrupole positions and field gradients were held fixed as well as the initial Twiss parameters of the electron beam. This means that optimal focusing of the electrons through the undulator line may not be achieved in all three cases; however, the transmission of the electron beam is sufficient to illustrate the variations in the polarizations of the optical field.

We consider SASE using a single, identical noise seed for each of these five choices for the ellipticity. This ensures that the injected electron phase space is also identical for each of these cases.

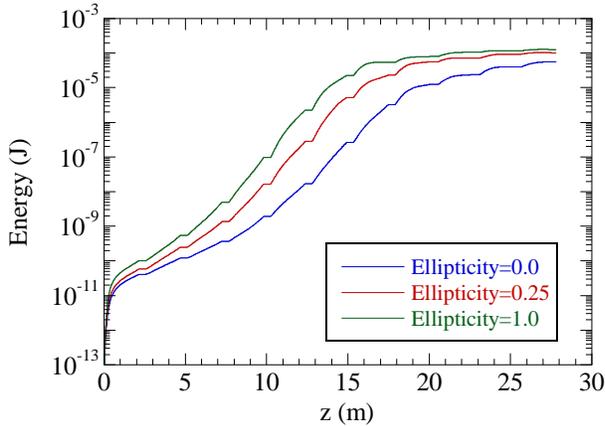

Fig. 7: Evolution of the energy vs $z$ for three choices of the ellipticity as seen in the new formulation.

The evolution of the ensemble average of the pulse energies versus distance along the undulator line is shown in Fig. 7 for three choices of the ellipticity: $u_e = 0.0$, 0.50, and 1.00. We observe that the distance to saturation increases and, hence, the growth rate decreases, as the ellipticity decreases from a value corresponding to pure linear polarization to one corresponding to circular polarization. This follows the decrease in the coupling coefficient associated with the increase in the oscillation in the magnitude of the transverse velocity. This is usually described by the so-called $JJ$-factor. As pointed out by J.R. Henderson et al. [27], the generalized $JJ$-factor is given by

$$JJ = \sqrt{1+u_e^2}\,\frac{K}{\sqrt{2}}\left[J_0(\zeta)-\frac{1-u_e^2}{1+u_e^2}J_1(\zeta)\right], \quad (21)$$

where

$$\zeta = \frac{\left(1-u_e^2\right)K^2/4}{1+\left(1+u_e^2\right)K^2/2}, \quad (22)$$

for an elliptic undulator. Observe that this reduces to the usual expressions in the limits of planar ($u_e = 0$) and helical ($u_e = 1$) undulators.

An important validation measure is a comparison between the exponentiation (gain) length ($L_G$) found in simulation with a theoretical estimate. A theoretical expression for the exponentiation length for elliptic undulators was given in [27] as a generalization of the parameterization described by Ming Xie [28] using the generalized $JJ$-factor shown in Eq. (21). Using this generalization, a comparison between the gain lengths found in simulation and theory for the five ellipticities under consideration is shown in Table 2 which shows good agreement.

| Ellipticity | $L_G$ (simulation) | $L_G$ (theory) |
|---|---|---|
| 0.00 | 0.80 m | 0.75 m |
| 0.25 | 0.79 m | 0.72 m |
| 0.50 | 0.69 m | 0.67 m |
| 0.75 | 0.60 m | 0.64 m |
| 1.00 | 0.58 m | 0.63 m |

Table 2: Comparison between the gain lengths found in simulation (new formulation) and theory.

As in the cases for the planar and helical undulators, energy conservation is maintained to within better than one part in $10^3$ for the APPLE-II undulator lines.

Using this present diagnostic for the Stokes parameters, the variation in the optical eccentricity versus the ellipticity of the undulator field is shown in Fig. 8 in blue while the prescribed polarization used in the fixed-phasor representation in the old formulation is indicated by the red line. Since the optical field is composed of multiple temporal slices, each of which may behave somewhat differently, the average ellipticity shown in the figure is a weighted average over the power in each slice. As is evident in the figure, the eccentricity is unity and $\theta$ corresponds to the direction of wiggle-motion when the Apple-II undulator is configured for linear polarization, as expected, and decreases approximately linearly with increasing ellipticity but more slowly than would be expected from the fixed-phasor representation. We do not find that the eccentricity is zero when the Apple-II undulator is configured for an ellipticity of unity which is close to that of a helical field. It is important to remark here that the Apple-II undulator model is not the same as the helical undulator field used in Sec. III.B and, in general, diverges from the helical field slowly away from the axis of symmetry. Thus, it may not be possible to achieve pure circular polarization with an Apple-II undulator. The seeded FEL at Fermi Light Source in Trieste uses APPLE-II undulators and, while this may be due in part to the optical beam transport line from the end of the undulator line, they have indeed observed that the optical polarization is not purely circular [13] at the location of measurement when the ellipticity of the undulators is set to unity.

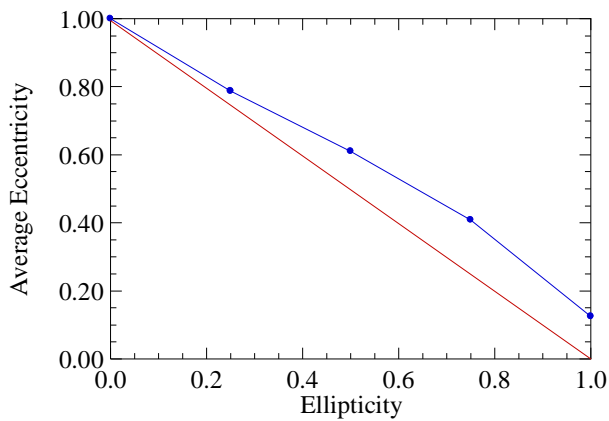

Fig. 8: The power-weighted average of the on-axis optical eccentricity for the Apple-II undulator line (blue line and markers). The prescribed eccentricity used in the old formulation with fixed phasor representation is shown in the red line.

As in the aforementioned cases, this undulator line presents a fixed polarization over the entire distance; hence, a fixed phasor representation of the optical field may be appropriate. In order to test this, we consider a single polarization corresponding to a helical undulator with an ellipticity of unity. In addition, we consider a single noise seed which generates identical initial phase spaces for both formulations.

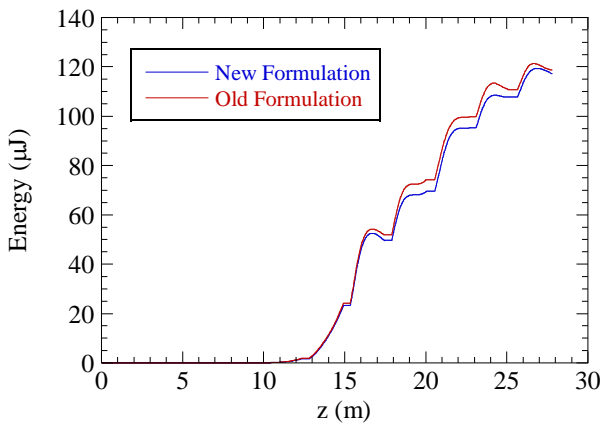

Fig. 9: Comparison of the old and new formulations for the APPLE-II undulator line with an ellipticity of unity.

A comparison plot showing the pulse energy over the course of the undulator line obtained from the old and new formulations is shown in Fig. 9. The rms discrepancy over the entire undulator line is only about 9%. In addition, the rms spot size found using the two formulations shows a discrepancy of only about 0.8%. The polarization used in the old formulation is purely circular and, while these discrepancies are not large, these comparisons reveal the limitations of a fixed phasor representation when simulating APPLE-II undulators which diverge from a purely helical undulator off the axis of symmetry.

## IV. SUMMARY AND CONCLUSION

Our purpose in this paper is twofold. First to present a numerical formalism that allows for the self-consistent evolution of the optical polarization when the electrons propagate through an arbitrary undulator line. This formalism not only allows for the simulation of arbitrary combinations/sequences of undulators with varying magnetic polarizations, but also for the simulation of how imperfections in the magnetic orientation affect the FEL interaction; particularly on the purity of the polarization state when appropriate models for such imperfections are implemented. This can also be extended to study the effects of magnet degradation that may occur over time for permanent magnet-based undulators. Second, to study the utility of the concept and the functionality of the formalism. To this end, we described validations of the formalism by comparison with the experimental results of the SPARC experiment which employed planar undulators and by comparison with an earlier formalism for a helical undulator configuration, and found excellent agreement for both of these cases. We found good to excellent agreement for various properties of the optical field, such as pulse energy, mode size and spectrum, when compared with the old formulation with prescribed optical polarizations. We also found that the simulations were in close agreement with the predicted growth rates and showed that energy conservation was maintained to within better than one part in $10^3$. However, we found that, with the new formulation, the polarization state starts to deviate from the fixed-phasor expectation when the ellipticity of the Apple-II undulator increases from zero to one, *i.e.*, the light is not circularly polarized when the ellipticity is unity.

We conclude that this formulation is able to faithfully simulate the control of the polarization of the output optical field for a variety of undulator combinations/sequences.

## ACKNOWLEDGEMENTS

This research used resources provided by the University of New Mexico Center for Advanced Research Computing, supported in part by the National Science Foundation.